\documentclass{ws-ijgmmp}

\newcommand{\vv}{``}

\newcommand{\mkk}{m_{_{\rm KK}}}

\begin{document}

\markboth{C. Branchina, V. Branchina, F. Contino and A. Pernace}
	{Dark Dimension and the Effective Field Theory limit}

%
\catchline{}{}{}{}{}
%

\title{DARK DIMENSION AND THE EFFECTIVE FIELD THEORY LIMIT}

\author{CARLO BRANCHINA$^{1\,2\,\dagger}$, VINCENZO BRANCHINA$^{3\,}$\footnote{Corresponding author}\,\,\,, FILIPPO CONTINO$^{3\,\ddagger}$, ARCANGELO PERNACE$^{3\,\star}$}

\address{
	${}^1$Department of Physics, Chung-Ang University, \\
	Seoul 06974, Korea \\
	${}^2$Department of Physics, University of Calabria, and INFN-Cosenza\\
	Arcavacata di Rende, I-87036, Cosenza, Italy \\
	${}^3$Department of Physics, University of Catania, and INFN-Catania\\
	Via Santa Sofia 64, I-95123 
	Catania, Italy\\
	\email{${}^\dagger$carlo.branchina@unical.it} 
	\email{${}^*$vincenzo.branchina@ct.infn.it} 
	\email{${}^\ddagger$filippo.contino@ct.infn.it}
	\email{${}^\star$arcangelo.pernace@ct.infn.it} 
}

\maketitle

\begin{history}
\received{(Day Month Year)}
\revised{(Day Month Year)}
\end{history}

\begin{abstract}
In\,\cite{Branchina:2023ogv} we pointed out that in the  Dark Dimension scenario\,\cite{Montero:2022prj} theoretical issues arise 
when the prediction for the vacuum energy $\rho$, that is obtained  from swampland conjectures in string theory, is confronted with the  corresponding result for $\rho$ in the effective field theory (EFT) limit. 
One of the problems concerns the widely spread belief that in higher dimensional EFTs with compact dimensions the vacuum energy is automatically finite. On the contrary, our analysis shows that $\rho$ contains (previously missed) UV-sensitive terms.  
Our work was challenged in\,\cite{Anchordoqui:2023laz}.  Here we show why in our opinion  the claims  in\,\cite{Anchordoqui:2023laz} are flawed, and provide further support to our findings. We conclude presenting ideas on the physical mechanism that should dispose of the large UV contributions to $\rho$.
\end{abstract}


\section{Introduction}

In search for a solution to the naturalness/hierarchy problem, in the nineties the possibility of having large extra dimensions (millimetre size) was considered\,\cite{Arkani-Hamed:1998jmv,Antoniadis:1998ig}. Recently, these ideas were applied to the so-called dark dimension (DD) scenario\,\cite{Montero:2022prj}, according to which we might well live in a $(4+1)$D universe with one compact dimension of micrometre size. In a nutshell\footnote{For  convenience of the reader less familiar with the DD proposal,  in Appendix A we provide a more detailed summary of the  different aspects and steps that lead to this scenario. }, the DD proposal\,\cite{Montero:2022prj}
combines swampland conjectures with observational data and phenomenological bounds, and is  based on the assumption that our universe lies in a unique (asymptotic) corner of the quantum gravity landscape, where only two kinds of light tower states  are expected: 
towers of string excitations 
or towers of KK states. Indicating with $\mu_{tow}$ the tower scale, and with $\Lambda_{cc}$ the measured vacuum energy (cosmological constant times $M_P^2$, with $M_P$ the Planck mass), the application of the distance conjecture to (A)dS vacua relates the two as
\begin{equation}\label{distcon1}
	\mu_{tow}\sim\left|\frac{\Lambda_{cc}}{M_P^4}\right|^{\alpha}M_P\,.
\end{equation} 
Furthermore, combining indications from  one-loop string calculations with the Higuchi bound\,\cite{Higuchi:1986py}, the proponents of the DD scenario 
restrict the parameter $\alpha$ to the range 
\begin{equation}\label{AdSconj1}
	\frac 14\leq \alpha\leq 
	\frac 12\,.
\end{equation} 
Then, considering bounds set by the non-observation of deviations from Newton's force law, they infer that  $\mu_{tow}$ is close to the energy scale associated to the cosmological
constant (that coincides with the neutrino scale)
\begin{equation}\label{alpha}
	\mu_{tow} \sim \Lambda_{cc}^{1/4}\sim \,\, {\rm meV}\,. 
\end{equation}  
They also observe that,  for\,\eqref{distcon1}  to be consistent with  \eqref{alpha},  it is necessary to take $\alpha$ at the lower bound of\,\eqref{AdSconj1},  $\alpha=1/4$.

The authors then make the following crucial observation. Let us quote from\,\cite{Montero:2022prj}: \vv since we can describe physics above the neutrino scale with  effective field theory'' (EFT), the string scenario is \vv ruled out experimentally''. They then conclude 
that the light tower must be a KK tower, $\mu_{tow}=m_{_{\rm KK}}$, so that\,\eqref{distcon1} becomes 
\begin{equation}\label{CC-rel1}
	\Lambda_{cc} 
	\sim m_{_{\rm KK}}^4\,.
\end{equation}
Finally, to  determine the number of extra compact dimensions, they compare bounds on $m_{_{\rm  KK}}$ that come from calculations of the  heating of neutron stars due to the surrounding cloud of trapped KK gravitons \cite{Arkani-Hamed:1998sfv}, \cite{Hannestad:2003yd}, \cite{PDG} with the value of $m_{_{\rm KK}}$  in \eqref{CC-rel1}. This allows them to conclude that there is   only one extra dimension of size $\sim \mu {\rm m}$.

In a recent paper\,\cite{Branchina:2023ogv}, we analysed aspects of
this scenario, and pointed out possible  theoretical issues. Below we briefly recall the scope and the main results of\,\cite{Branchina:2023ogv}. As already said, according to the proponents of the  DD scenario,  above the $m_{_{\rm KK}}$ scale our universe is described by a $(4+1)$D EFT with one compact dimension of micrometer size. This  proposal has  triggered a large amount of work, where  $(4+1)$D EFT models that could  realize the DD scenario are proposed, and their phenomenological consequences studied (see for instance\,\cite{Anchordoqui:2022txe,Greene:2022urm,Gonzalo:2022jac,Anchordoqui:2022tgp,Anchordoqui:2022svl,Anchordoqui:2023oqm,Anchordoqui:2023wkm,Anchordoqui:2023qxv,Cui:2023wzo,Obied:2023clp,BitaghsirFadafan:2023hwg}). In\,\cite{Branchina:2023ogv}, our goal  was to investigate on the matching between the string theory framework in which the basic relation\,\eqref{CC-rel1} of the DD proposal is obtained and its low energy EFT limit. 

To this end, we  considered a supersymmetric (4+1)D theory with Scherk-Schwarz SUSY breaking (and also a non-supersymmetric theory) with  compact dimension in the shape of a circle\footnote{In the DD scenario the Standard Model (SM) lives on a 3-brane, and it would  be then more appropriate to consider an orbifold. For our scope,  namely the calculation of the one-loop contribution of bulk fields to the vacuum energy, it is sufficient to consider the simplest case of a circle.} of radius $R$ 
and calculated the one-loop vacuum energy $\rho^{1l}$. According to existing literature (see for instance\,\cite{Kantowski:1986iv,Ponton:2001hq,Ito:2003tc,Ghilencea:2005vm}, and for the analogous calculation of the effective potential \cite{Antoniadis:1998sd, Delgado:1998qr}),  in $(4+n)$D theories with $n$ compact dimensions  $\rho^{1l}$ should be finite and proportional to $m_{_{\rm KK}}^4$,
\begin{equation}\label{1l-usual}
	\rho^{1l} \sim m_{_{\rm KK}}^4\,,
\end{equation}
with no need for fine-tuning.  In\,\cite{Branchina:2023ogv} we showed that the right hand side of\,\eqref{1l-usual} also contains UV-sensitive terms (see the first line in the right hand side of Eq.\,\eqref{SUSYtot*} below), that were  missed in previous literature\footnote{The presence of similar UV-sensitive terms in the calculation of the effective potential of a scalar theory was first hinted in\,\cite{Ghilencea:2001ug}. Moreover, it was shown in\,\cite{Ghilencea:2001bw} that the presence of Fayet-Iliopulos terms induces further divergences.}.
Therefore, the matching between the prediction\,\eqref{CC-rel1} for the vacuum energy $\Lambda_{cc}$, that comes from swampland conjectures in string theory, and the corresponding result in the low energy field theory limit is  not as straightforward as it was thought on the basis of  the   result\,\eqref{1l-usual}. 
In this respect, it is worth to stress again (see the paragraph below Eq.\,\eqref{alpha}) that such a matching is crucial for the DD scenario. 

In\,\cite{Branchina:2023ogv} we show why it is commonly thought  that $\rho^{1l}$ 
is finite and proportional to $m_{_{\rm KK}}^4$: this result comes from a mistreatment of the  loop momentum asymptotics. 
Performing the calculation properly, we found that (in addition to the finite term $\sim m^4_{_{\rm KK}}$ and to divergences that  cancel out in SUSY theories) the one-loop contribution to the vacuum energy  $\rho^{ 1l}$  contains \vv UV-sensitive'' terms proportional to powers of the boundary charges. These divergences {\it do not} disappear, not even in a SUSY theory.
This is due to the fact that, to trigger the Scherk-Schwarz mechanism, the boundary charges $q_b$ and $q_f$ of the boson and fermion superpartners need to be different.

Our work generated a debate that finally resulted in\,\cite{Anchordoqui:2023laz}, where our results were challenged. Below we present a deep analysis of the arguments given in\,\cite{Anchordoqui:2023laz}, and show why in our opinion these authors (in their attempt to confute our results) bring confusion on the whole subject using flawed arguments that lead them to incorrect conclusions. In view of the great  wealth of theoretical and phenomenological work triggered by the DD scenario, we believe it worth trying to bring clarity on the subject and continuing its investigation.  In the course of this analysis, we will bring new arguments that further support the conclusions of our previous work.

The rest of the paper is organised as follows. In section 2, we deeply investigate the arguments brought by the authors of\,\cite{Anchordoqui:2023laz} against our work, showing that they misuse our results and draw incorrect conclusions. Moreover, we present the global framework in which the DD scenario is framed, that allows to better understand the flaws contained in the claims made in\,\cite{Anchordoqui:2023laz}. In section 3, we show that arguments typically used in the literature (including\,\cite{Anchordoqui:2023laz}) to support the widely spread belief that in higher dimensional theories with compact dimensions the vacuum energy is automatically finite are misleading. In section 4, we present our conclusions, together with ideas on the physical mechanism that in our opinion should dispose of the large UV contributions to $\rho^{1l}$.

\section{Global framework of the DD scenario and misuse of our results}

\begin{figure*}
	\centering
	\includegraphics[scale=0.65]{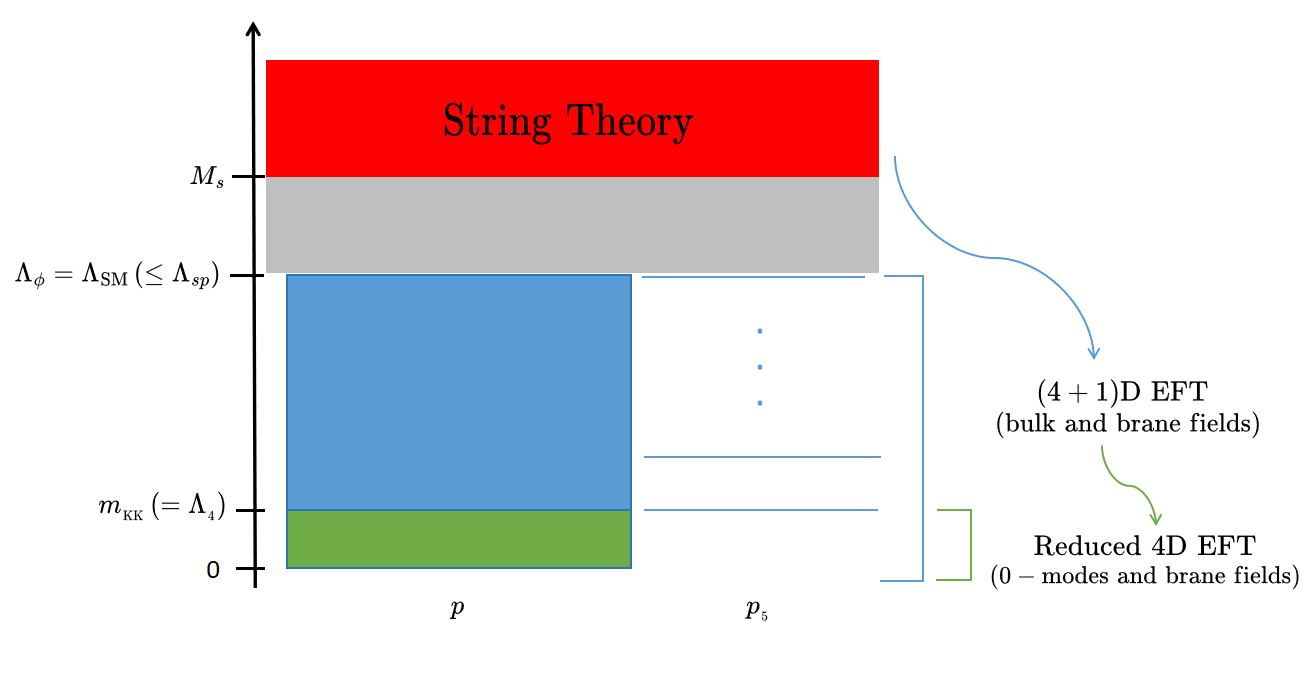}
	\caption{Pictorial representation of the energy scales and theories involved in the Dark Dimension scenario. The grey region indicates a possible UV-completion of the $(4+1)$D theory between its (rescaled, see text) physical cutoff $\Lambda_\phi$ and the string scale $M_s$. A particular case could be that $\Lambda_\phi$ directly coincides with $M_s$. Below $\Lambda_\phi$ (that coincides with the Standard Model physical cutoff $\Lambda_{\rm SM}$, see text) the $(4+1)$D theory of the DD scenario (blue region), that contains bulk fields and SM (with possibly other) brane fields, takes place. Below $m_{_{\rm KK}}$, the theory is well described by a $4D$ EFT (green region) containing the $0$-modes of the bulk fields together with brane fields. The components $p$ and $p_5$ of  the  $(4+1)$D momentum $\hat p\equiv (p,p_{_5})$, with $p\equiv (p_1,p_2,p_3,p_4)$ and $p_5$ the fifth component, are reported. The fully colored blue and green rectangles indicate that $p$ takes continuous values (non-compact dimensions), while the blue lines indicate that $p_5$ takes discrete values starting at $m_{_{\rm KK}}$ (compact dimension).}
	\label{fig}
\end{figure*}

As anticipated above, our results were challenged by the authors of\,\cite{Anchordoqui:2023laz}. Below we  show that they misuse and distort our findings, and how this leads them to  incorrect conclusions. Moreover, we present the global framework in which the DD scenario is framed (see Fig.\,\ref{fig}), which further helps in bringing clarity on the whole subject.
Before doing that, however, it is worth mentioning that\,\cite{Anchordoqui:2023laz} starts with the claim that we question the swampland relation\,\eqref{CC-rel1}, while what we did in\,\cite{Branchina:2023ogv} is totally different, and this was far from being our goal (see the Introduction above). 

Our results for  $\rho^{ 1l}$ are  in Eqs.\,(25) and (26)  of\,\cite{Branchina:2023ogv}. For ease of presentation (and with no loss of generality), from now on  we concentrate on the SUSY theory case only (unless expressly indicated otherwise), and write the  contribution to  $\rho^{1l}$  from a single couple of superpartners for the massless case (i.e.\,taking $m=0$ in (25) and (26) of\,\cite{Branchina:2023ogv})

\begin{align}\label{SUSYtot*}	\rho^{1l}&=\frac{(q^2_b-q^2_f)\, R\,\mkk^2}{60\pi^2}\Lambda^3-\frac{(q_b^4-q_f^4)\,R^{\frac 13}\,\mkk^{\frac{10}{3}}}{280\pi^2}\Lambda\nonumber\\
	&+\frac{\,3\left[\text{Li}_5(e^{2 \pi  i q_b})-\text{Li}_5(e^{2 \pi  i q_f})+h.c.\right]\,}{128 \pi ^6}\,\mkk^4\,,
\end{align}
where $q_b$ and $q_f$ are the boson and fermion boundary charges  respectively, and $\text{Li}_5(x)$ is the polylogarithm function $\text{Li}_n(x)$ for $n=5$.  $\Lambda$ is the physical  UV cutoff of the  $(4+1)$D theory 
defined by the action  
$\mathcal{S}^{(4+1)}=\mathcal{S}_{\rm grav}^{(4+1)}+\mathcal{S}_{\rm{matter}}^{(4+1)}$,  
with $(4+1)$D metric $\hat g_{_{MN}}$  parametrized in terms of the graviphoton $A_\mu$ and the radion 
field $\phi$ (see Appendix B, and  equations (6)-(12) in\,\cite{Branchina:2023ogv}). The KK scale $m_{_{\rm KK}}$ is given in terms of the radion field {\it vev} (here indicated with $\phi$)
\begin{equation}\label{emmekk}
	m_{_{\rm KK}}=e^{\sqrt{\frac32}\frac{\phi}{M_P}}R^{-1}\equiv R_\phi^{-1}\,.
\end{equation}

To be phenomenologically viable, the $(4+1)$D model that implements the DD scenario has to include the SM on a 3-brane. An important consequence is that the physical cutoff $\Lambda$ of the $(4+1)$D theory is an upper bound for the SM  cutoff $\Lambda_{\rm SM}$ (the derivation of Eq.\,\eqref{SM-cut} below is in Appendix B)  
\begin{equation}\label{SM-cut}
	\Lambda_{\rm SM} \leq \Lambda\,.
\end{equation}
As we will see, this result is important for our analysis. Let us also note that  $\mkk$ is the cutoff of the \vv reduced 4D theory'', which besides brane fields contains the zero modes of bulk fields. Moreover, it is worth to  observe that, although both the Standard Model and this \vv reduced  theory'' are $4$-dimensional EFTs,    they are certainly not the same theory, and their cutoffs ($\Lambda_{\rm SM}$ and $\mkk$ respectively) should not be confused. In Fig.\,\ref{fig} we provide  a  comprehensive picture of the global framework in which the DD scenario is framed, described in detail in the caption below. 

In Eq.\,\eqref{SUSYtot*}, the  divergences (i.e.\,UV-sensitive terms)  
are in the first line, the finite term in the second one.  Limiting ourselves to consider  only the {\it most divergent} term of $\rho^{1l}$ (and doing the same for the non-SUSY theory, whose  result for $\rho^{1l}$ is not reported above but can be found in our work\,\cite{Branchina:2023ogv}), we have
\begin{equation}\label{domrho**} 
	\rho^{1l}  \sim m_{_{\rm KK}}^2 R \Lambda^3 \qquad {\rm and} \qquad \rho^{1l}  \sim m_{_{\rm KK}}^{2/3} R^{5/3} \Lambda^5\,,
\end{equation}
for the SUSY and the non-SUSY case respectively.
These latter equations  are Eqs.\,(28) and (29) of\,\cite{Branchina:2023ogv}.  As stressed above, one of the main goals of our work\,\cite{Branchina:2023ogv}  is to point out that  $\rho^{1l}$ contains previously unnoticed divergences (that do not disappear  even in a SUSY theory)  and that, before comparing  $\rho^{1l}$ with the measured value of the vacuum energy $\Lambda_{cc}$, a fine-tuning is needed. 

For the purposes of the present analysis, it is  of fundamental  importance to stress  that  the authors of\,\,\cite{Anchordoqui:2023laz} write our Eqs.\,\eqref{domrho**} as
(see Eqs.\,(4) of\,\cite{Anchordoqui:2023laz})
\begin{equation}\label{domlambda*} 
	\Lambda_{cc}  \sim m_{_{\rm KK}}^2 R \Lambda^3 \qquad {\rm and} \qquad \Lambda_{cc}  \sim m_{_{\rm KK}}^{2/3} R^{5/3} \Lambda^5\,, 
\end{equation}
thus operating in\,\eqref{domrho**} the replacement  
\begin{equation}\label{replace*}
	\rho^{1l} \to \Lambda_{cc}\,.
\end{equation} 
So doing,  
they take the result $\rho^{1l}$ of our one-loop calculation in\,\cite{Branchina:2023ogv}  to {\it directly coincide}  (without any fine-tuning) with the measured value of the vacuum energy.

Although this is opposite to what we did (we repeat ourselves, in\,\cite{Branchina:2023ogv} our goal was to point out the presence of unexpected UV-sensitive terms), the replacement\,\eqref{replace*} is not in itself a problem. It is a theoretically legitimate possibility that the cutoff $\Lambda$ is a physical scale that appears in physical quantities,  with no need for  fine-tuning. However, the consequences of such a choice  have to be fully taken into account. 
Anticipating on the results, we will show that the replacement $\rho^{1l} \to \Lambda_{cc}$  has crucial physical
consequences that the authors of\,\cite{Anchordoqui:2023laz} 
failed to recognize. It is precisely this replacement that leads them to incorrect conclusions. Let us note that, for reasons that will be soon apparent, the right hand side  of\,\eqref{domlambda*}$_1$  should contain all the terms that appear in the right hand side of\,\eqref{SUSYtot*}, not only the first one. However, since this has no impact on the present  discussion, we can equivalently refer to\,\eqref{domlambda*}$_1$ or more completely to\,\eqref{SUSYtot*}.

It is now crucial to realise that, once  $\rho^{1l}$ is chosen to coincide with the  measured vacuum energy $\Lambda_{cc}$, Eqs.\,\eqref{domlambda*} {\it completely determine} $\Lambda$.  As we will see, the authors of\,\cite{Anchordoqui:2023laz} 
base their analysis on  Eqs.\,\eqref{domlambda*} and, failing to realise that they do not have this freedom, freely play with the cutoff  
$\Lambda$.  
Below we determine the value of $\Lambda$ imposed by their choice\,\eqref{replace*}, 
and investigate on the possibility that the resulting  $\Lambda$ can really play the role of the theory UV cutoff. For this to be true,  $\Lambda$ cannot violate well-tested phenomenology, in particular it should not enter in conflict with the Standard Model. 

Let us determine now the value of $\Lambda$ imposed by\,\eqref{replace*}. First of all, we recall that from\,\eqref{CC-rel1} we have $m_{_{\rm KK}}\sim \Lambda_{cc}^{1/4}$. A simple inspection of \,\eqref{domlambda*}$_1$  (or\,\eqref{SUSYtot*} with the replacement\,\eqref{replace*}) shows that this implies\footnote{The same is true for\,\eqref{domlambda*}$_2$.} 
\begin{equation}\label{necessary}
	R\Lambda^3 \sim m^2_{_{\rm KK}}\sim ({\rm meV})^2\,. 
\end{equation}

The above relation is the {\it unavoidable} consequence of\,\,\eqref{replace*}.  
Starting from\,\eqref{necessary}, it is now easy to show that  the whole construction put forward in\,\cite{Anchordoqui:2023laz} is flawed. 
In fact, using the relation ($\hat M_P$ is the five-dimensional Planck scale)
\begin{equation}\label{MPhigh}
	\hat M_P^3=(2\pi R)^{-1} M_P^2\,,
\end{equation}
Eq.\,\eqref{necessary} is straightforwardly rewritten as 
\begin{align}
	\label{Lambda hatM}
	\Lambda^3  
	\sim \left(\frac{m^2_{_{\rm KK}}}{M_P^2}\right)\hat M_P^3,
\end{align}
from which we  obtain (remember that $m_{_{\rm KK}}\sim$ meV)
\begin{equation}
	\label{boundLambda}
	\Lambda \,\sim\, 
	10^{-21}\hat M_P \,\le\, 10^{-2}\, \text{GeV},
\end{equation}
where the last inequality stems from $\hat M_P\le M_P$. Eq.\eqref{boundLambda} seems to set a too low upper bound on the cutoff $\Lambda$ of the $(4+1)$D theory. We will see in a moment that this is really the case. 

The problem  comes from the relation between  $\Lambda$ and the $4$D cutoff $\Lambda_{\rm SM}$ of the Standard Model, that, as already said, lives on a 3-brane. In Appendix B (to which we refer for the calculation) we show that $\Lambda_{\rm SM}$ is  given by
\begin{equation}
	\label{LambdaSM}
	\Lambda_{\rm SM}=\Lambda_\phi\equiv e^{\frac{1}{\sqrt{6}}\frac{\phi}{M_P}}\Lambda,
\end{equation}
where $\phi$ is the $vev$ of the radion field (see comments above Eq.\,\eqref{emmekk} and   Appendix B) and $\Lambda_\phi$ is the $(4+1)$D cutoff for the momenta rescaled by the factor $e^{\frac{1}{\sqrt{6}}\frac{\phi}{M_P}}$, that are nothing but the momenta from the $4$D observer perspective (see Eq.\,\eqref{cutoff4} in Appendix B). Since the asymptotic scenario of the DD proposal is realized 
for negative values of $\phi$ (see Appendix A), Eqs.\,\eqref{boundLambda} and\,\eqref{LambdaSM} imply 
\begin{equation}\label{Lambdamin}
	\Lambda_{\rm SM}< 10^{-2}\, {\rm GeV}\,,
\end{equation}
an unacceptably low upper bound on $\Lambda_{\rm SM}$.

We can go further and determine the value that $\Lambda_{\rm SM}$ should have if the replacement $\rho^{1l} \to \Lambda_{cc}$ operated by the authors of\,\cite{Anchordoqui:2023laz} could be accepted. In fact, observing that the species scale $\Lambda_{\rm sp}$\,\cite{Dvali:2007hz,Dvali:2007wp} is given by
\begin{equation}\label{Lambdaspec} 
	\Lambda_{\rm sp}\sim m_{_{\rm KK}}^{1/3}M_P^{2/3}\sim e^{\frac{1}{\sqrt{6}}\frac{\phi}{M_P}}\hat M_P 
\end{equation}
(see for instance\,\cite{Grimm:2018ohb} and the Appendix of\,\cite{Branchina:2023ogv}), comparing\,\eqref{LambdaSM} with \eqref{Lambdaspec} and using\,\eqref{boundLambda} we immediately have
\begin{equation}
	\label{Lambda SM Lambda sp rel}
	\Lambda_{\rm SM}\sim \frac{\Lambda}{\hat M_P}\Lambda_{\rm sp} \sim 10^{-21} \Lambda_{\rm sp}. 
\end{equation}
Finally, since from\,\eqref{Lambdaspec} we have $\Lambda_{\rm sp}\sim 10^9$ GeV,
\begin{equation}
	\label{Lambda SM iden}
	\Lambda_\phi=	\Lambda_{\rm SM} \sim 10^{-12} {\rm GeV} = 10^{-3} {\rm eV} \,,
\end{equation}
a too low value, as already known from\,\eqref{Lambdamin}. 

Eq.\,\eqref{Lambda SM iden} definitely shows that what the authors of \cite{Anchordoqui:2023laz} do when they present our  Eqs.\,\eqref{domrho**} as Eqs.\,\eqref{domlambda*} (that are nothing but \eqref{domrho**} with the replacement $\rho^{1l}\to \Lambda_{cc}$),
is physically not allowed. They operate an inconsistent replacement, and Eqs.\,\eqref{domlambda*} are deprived of physical meaning. 
The point is that   the arguments they use in their attempt to confute our work completely depend on this replacement, actually cannot even be formulated without it. Their arguments are thus unfounded and inconsistent. 

There is a simpler  way to obtain\,\eqref{Lambda SM iden} for $\Lambda_{\rm SM}$. From\,\eqref{emmekk} and\,\eqref{LambdaSM} it is easily seen that $R\Lambda^3 = R_\phi\Lambda_\phi^3$, so that Eqs.\,\eqref{domlambda*} can be written as
\begin{equation}\label{lambda***} 
	\Lambda_{cc}  \sim m_{_{\rm KK}}^2 R_\phi \Lambda_\phi^3 \qquad {\rm and} \qquad \Lambda_{cc}  \sim m_{_{\rm KK}}^{2/3} R_\phi^{5/3} \Lambda_\phi^5\,. 
\end{equation}
Therefore, inserting\,  $R_\phi=m_{_{\rm KK}}^{-1}$ in\,\eqref{lambda***},  we immediately find that $\Lambda_{cc}\sim m_{_{\rm KK}}^4$ is obtained {\it only} if 
\begin{equation}\label{neww}
	\Lambda_\phi=\Lambda_{\rm SM}\sim m_{_{\rm KK}} \sim 10^{-3}\,{\rm eV}\,,
\end{equation}
that is nothing but\,\eqref{Lambda SM iden}. Eq.\,\eqref{neww} (and equivalently Eq.\,\eqref{Lambda SM iden}) shows a further drawback of the (illegitimate) replacement $\rho^{1l}\to \Lambda_{cc}$. From this equation in fact we see that the latter  forces the $(4+1)$D cutoff $\Lambda_\phi$ to be  squeezed upon $m_{_{\rm KK}}$. This leaves no space to the $(4+1)$D EFT needed for the DD scenario, and one is only left with a $4$D theory valid below  $m_{_{\rm KK}}$ ($\sim$ meV), above which the theory directly merges to its UV completion (see Fig.\,\ref{fig}, where the choice $\Lambda_\phi=m_{_{\rm KK}}$ makes the blue region disappear). 
Therefore, if the   replacement $\rho^{1l}\to \Lambda_{cc}$ operated in\,\cite{Anchordoqui:2023laz} could be accepted, this would close the door to the DD scenario. Certainly, this was not the original intention of the authors in\,\cite{Anchordoqui:2023laz}. 

The above analysis shows that Eqs.\,\eqref{domlambda*} are physically inconsistent, and obviously no use can be made of them. In particular,  they cannot be used  to confute our results. Unfortunately, this is precisely what the authors of\,\cite{Anchordoqui:2023laz} do, thus developing untenable arguments and claims. 

The present section could end here. 
Nevertheless, in the hope that this can bring  further clarity and help  the reader gain more insight in the whole subject, we now provide a  detailed analysis of their  arguments, that (as already said) are  based on the inconsistent Eqs.\,\eqref{domlambda*}.  
To follow  their presentation  closely, below we consider both the SUSY and the non-SUSY cases, i.e.\,both their  Eqs.\,\eqref{domlambda*}$_1$ and\,\eqref{domlambda*}$_2$.  

One of the authors' first points is to compare the (A)dS distance conjecture\,\eqref{distcon1}, that for  the reader's convenience we rewrite here as ($\mu_{tow}$ in\,\eqref{distcon1} is clearly $m_{_{\rm KK}}$)
\begin{equation}\label{adistcon**}
	m_{_{\rm KK}}\sim M_P^{1-4\alpha}\,\Lambda_{cc}^{\alpha}\,,
\end{equation} 
with Eqs.\,\eqref{domlambda*}, that again for convenience we write as
\begin{equation}\label{domlambda2}
	m_{_{\rm KK}}\sim (R\Lambda^3)^{-1/2} 
	\Lambda_{cc}^{1/2} \,\,\,;\,\,\, m_{_{\rm KK}}\sim (R\Lambda^3)^{-5/2} 
	\Lambda_{cc}^{3/2}\,.
\end{equation}
This comparison leads them to affirm that  for the parameter $\alpha$ {\it we claim} the values $\alpha=1/2$ and $\alpha=3/2$,  for the SUSY and non-SUSY case respectively.

Nothing could be further from us. Such, allegedly  ours, claim actually results from (i) the replacement\,\eqref{replace*} that the authors of\,\cite{Anchordoqui:2023laz} operate 
in our  equations\,\eqref{domrho**}, thus obtaining Eqs.\,\eqref{domlambda*},  and (ii) the comparison of these latter equations  with \eqref{adistcon**}. Since we know that   Eqs.\,\eqref{domlambda*} cannot be written (simply do not exist), we never  could have made such a claim. 
Having clarified this point, we further observe that even if for a moment we follow the authors' line of reasoning, it is immediate to see from\,\eqref{adistcon**} and\,\,\eqref{domlambda2} 
that 
the values $\alpha=1/2$ and $\alpha=3/2$ (for the SUSY and non-SUSY case, respectively) can be derived only if it is also verified that
\begin{equation}\label{necessa}
	R\Lambda^3 \sim M_P^2\,.
\end{equation}
Closing for a moment our eyes on the phenomenological impossibility of having\,\eqref{Lambda SM iden}, what we have just found would mean that, for\,\eqref{distcon1}  and\,\,\eqref{domlambda*} to give $\alpha=1/2$ and $\alpha=3/2$  (for the SUSY and non-SUSY case, respectively), Eqs.\,\eqref{necessary} and\,\eqref{necessa} should both be  fulfilled at the same time, which is obviously an absurd requirement. 

Let us move now to another of the arguments developed in\,\cite{Anchordoqui:2023laz}. The authors   consider the Higuchi bound\,\footnote{We remind here that according to the Higuchi bound the {\it physical} mass $M$ of a spin-2 particle in dS space and the {\it physical} value of the cosmological constant satisfy\,\cite{Higuchi:1986py}, 
	\begin{equation}
		M^2\ge \frac{2}{3}\frac{\Lambda_{cc}}{M_p^2}\,. 
\end{equation}},
that in the present case reads
\begin{equation}
	\label{Higuchi***}
	\Lambda_{cc}
	\le \left( \frac{3}{2}M_P^2\right) m_{_{\rm KK}}^2\,,
\end{equation}
and comparing \eqref{Higuchi***} with \eqref{domlambda*} they get 
\begin{equation}\label{cut1*}
	R\Lambda^3 \lesssim M_P^2 \qquad {\rm and} \qquad 
	R\Lambda^3 \lesssim 10^{-48} M_P^2
\end{equation}
for the SUSY and non-SUSY cases, respectively\footnote{To obtain \eqref{cut1*}$_2$,  
	$\Lambda_{cc}\sim 10^{-120} M_P^4$ is used.}. 
As we know,  Eqs.\,\eqref{cut1*} would not even exist without the inconsistent replacement $\rho^{1l} \to \Lambda_{cc}$ in our equations\,\eqref{domrho**}. Nevertheless, closing again an eye on this point, we now follow  their reasoning.  Concerning \eqref{cut1*}$_1$, they observe  that this relation does not constrain  $R\Lambda^3$ too  strongly, while for \eqref{cut1*}$_2$ they say it imposes a too strong constraint on  $\Lambda$. This latter claim  deserves a comment. Using the relation\,\eqref{MPhigh}, Eq.\,\eqref{cut1*}$_2$ can be written as 
\begin{equation}\label{mild}
	\Lambda \lesssim 10^{-16} \, \hat M_P\,.
\end{equation}
At the same time, 
using again $R\Lambda^3=R_\phi\Lambda_\phi^3$ (see comments above Eq.\eqref{lambda***}), Eq.\,\eqref{cut1*}$_2$  can also be written in terms of  $\Lambda_\phi$ as
\begin{equation}\label{toolow}
	\Lambda_{\rm SM}=\Lambda_\phi\lesssim 10^5\,m_{_{\rm KK}}\sim 10^2 \,{\rm eV}.
\end{equation}

A simple look at \eqref{mild} and\,\eqref{toolow} shows that the trouble with\,\eqref{cut1*}$_2$ is not that it implies a too low value for the cutoff $\Lambda$ (as stated in\,\cite{Anchordoqui:2023laz}), but rather that it implies a too low value for the SM cutoff $\Lambda_{\rm SM}$, as we see from\,\eqref{toolow}. Yet another contradiction is found if we observe that,   combining\,\eqref{LambdaSM} with\,\eqref{toolow},  at threshold (i.e.\,when in\,\eqref{toolow} $\Lambda_\phi\sim 10^2$ eV) we have  
\begin{equation}\label{toolowbis}
	\Lambda \gtrsim 10^2 \,{\rm eV}. 
\end{equation}
In fact, few lines after their claim that\,\eqref{cut1*}$_2$ implies a too low value for $\Lambda$, the authors of \cite{Anchordoqui:2023laz}
make another claim, namely that the \vv correct choice'' for $\Lambda$ is $\Lambda=m_{_{\rm KK}}$ (see below). As we have just seen, however, at threshold  \eqref{cut1*}$_2$ implies \eqref{toolowbis}, while $\Lambda=m_{_{\rm KK}}$ means $\Lambda \sim$ meV, that is five orders of magnitude below the lower bound in\,\eqref{toolowbis}. 
The patent  contradiction between these two claims of theirs can hardly be missed. 

Successively, in\,\cite{Anchordoqui:2023laz}  the authors make yet another use of their Eqs.\,\eqref{domlambda*}.  
They observe that these equations still contain the radius $R$, and claim that the latter should be replaced by $m_{_{\rm KK}}^{-1}$, so that additional powers of $m_{_{\rm KK}}$ appear in these equations. Before proceeding with our analysis, we must observe that, so doing, the authors miss the crucial radion dependence  in the relation between $R$ and $m_{_{\rm KK}}$ (see\,\eqref{emmekk}). 

Anyway, let us go on and follow their  arguments.
Invoking the UV-IR mixing occurring  in string theory and/or quantum gravity, the authors of\,\cite{Anchordoqui:2023laz} say that the UV cutoff $\Lambda$ of an  EFT coupled to quantum gravity should depend on the IR scale, that in this case is  $m_{_{\rm KK}}$. They then observe that the largest possible cutoff is the species
scale $\Lambda_{\rm sp}\sim m_{_{\rm KK}}^{1/3}M_P^{2/3}$ (see \eqref{Lambdaspec}), which in fact depends on $m_{_{\rm KK}}$. Now, 
replacing in   Eqs.\,\eqref{domlambda*}  $R$ with $m_{_{\rm KK}}^{-1}$ and  $\Lambda$ with $\Lambda_{\rm sp}$, they finally get\footnote {Actually $\Lambda_{\rm sp}$ is a cutoff for the 4D momentum $p$ and for the KK tower, (see for instance\,\cite{Grimm:2018ohb} and the Appendix of\,\cite{Branchina:2023ogv}), while $\Lambda$ is the cutoff for the $(4+1)$D  momentum $\hat p$, that gives rise to the cutoff  $\Lambda_\phi$ for $p$, see Eq.\,\eqref{cutoff4} in Appendix B. Therefore, contrary to what is done in\,\cite{Anchordoqui:2023laz}, $\Lambda_{\rm sp}$ has to be identified with $\Lambda_\phi$ not with  $\Lambda$. 
	However, since  $R\Lambda^3 = R_\phi\Lambda_\phi^3$ (see comments above\,\eqref{lambda***}), the two mistakes made by the authors in replacing $R$ (rather than $R_\phi$) with $m_{_{\rm KK}}^{-1}$ and $\Lambda$ (rather than $\Lambda_\phi$) with $\Lambda_{\rm sp}$ cancel between one another. Therefore, the correct substitutions lead to the same result\,\eqref{lambda**}.
	Moreover, we note that $\Lambda_{\rm sp}$ is the maximal value for  $\Lambda_\phi$, and is obtained for $\Lambda=\hat M_P$, which is of course the maximal possible value for $\Lambda$.} 
\begin{equation}\label{lambda**} \Lambda_{cc}  \sim m_{_{\rm KK}}^2 M_P^2 \qquad {\rm and} \qquad \Lambda_{cc}  \sim m_{_{\rm KK}}^{2/3}  M_P^{10/3}\,. 
\end{equation}
From these results, they conclude that this first of their  \vv UV-IR mixing motivated'' attempts is not satisfactory, as\,\eqref{lambda**}$_2$ violates the Higuchi bound.

Then, still invoking the \vv UV-IR mixing'', they proceed by making another  trial for $\Lambda$, that they eventually dub as the \vv correct choice''. In this second attempt they insert in Eqs.\,\eqref{domlambda*}   $\Lambda=m_{_{\rm KK}}$   (and $R=m_{_{\rm KK}}^{-1}$, as for the previous case). A first thing worth to note in connection with this new \vv trial'' is that  a simple look to Eqs.\,\eqref{domlambda*} shows that, since  they are fixing all the scales to  $m_{_{\rm KK}}$, the right hand side of these equations is trivially proportional to $m_{_{\rm KK}}^4$. The authors of\,\cite{Anchordoqui:2023laz} seem to be satisfied by this result, and claim that with this \vv correct choice'' of $\Lambda$ they are able to remove what they call the \vv ambiguity'' in the EFT calculation and  fix what they consider as the unique cutoff required by UV-IR mixing.  In our opinion, this is (at least) too cheap a way to get the result \vv vacuum energy $\sim m_{_{\rm KK}}^4$'', and to implement the string UV-IR connection in the field theory limit. But this is not the worst. What certainly kills the possibility of taking  $\Lambda=m_{_{\rm KK}}$ is what we have already seen. This choice is not allowed as it implies a ridiculously small cutoff for the Standard Model\footnote{It is probably worth to note that if instead of the authors' replacement 
	$R=m_{_{\rm KK}}^{-1}$  one makes the correct replacement  $R_\phi=m_{_{\rm KK}}^{-1}$ (see\,\eqref{emmekk}), from\,\eqref{lambda**} we have that $\Lambda_{cc}\sim m_{_{\rm KK}}^4$ is obtained for $\Lambda_\phi=m_{_{\rm KK}}$ rather than
	$\Lambda=m_{_{\rm KK}}$. In any case, both $\Lambda_\phi=m_{_{\rm KK}}$ and
	$\Lambda=m_{_{\rm KK}}$  are unacceptably low.}.
	
\section{Casimir energy and finite temperature field theory}

We finally comment on two additional arguments that the authors of\,\cite{Anchordoqui:2023laz} use to lend further support to their views\footnote{ These arguments are not new, and have been repeatedly used over the years as they seem to substantiate the widely spread belief that for field theories with compact extra dimensions the vacuum energy turns out to be automatically finite.}. Referring to 
the finiteness of string theory calculations, their goal is to give   examples where finite results are obtained also  in field theory. To this end, they consider (i) the
calculation of the one-loop free energy $F^{1l}$ in finite temperature QFT, where the $T$-dependent contribution  goes like $F_T^{1l}\sim T^4 $ with no need for any  \vv string regularization'' (to use their words), and (ii) the calculation of the Casimir energy.

Let us begin with the finite temperature field theory case.
As stressed in\,\cite{Branchina:2023rgi}, despite their apparent similarity, there is a profound difference between the sum over the integer $n$ that appears in  $F^{1l}$ and the analogous sum in the calculation of the vacuum energy $\rho^{1l}$ in KK theories. In fact,  both for   $\rho^{1l}$ and $F^{1l}$ we have an expression of the kind (below $d=4$ for the $(4+1)$D KK theory and $d=3$ for finite temperature field theory)
\begin{equation}\label{common}
	\rho^{1l}\,\, , \,\,  F^{1l} \, \sim \frac12\,\sum_n\int\,d^d p \, \log(p^2+ m^2+ f_n)\,.
\end{equation}

For the case of KK theories ($\rho^{1l}$) the integer $n$ is related to the fifth component of the $(4+1)$D loop momentum $\hat p$, that is cut at $\Lambda$. Therefore $n$ and $p$ are intertwined between one another and are to be cut accordingly\,\cite{Branchina:2023ogv,Branchina:2023rgi}. This in particular implies an upper bound on the integer $n$. 
As a result, our Eqs.\,(25) and (26)  of\,\cite{Branchina:2023ogv}, here reported for the massless SUSY case in Eq.\,\eqref{SUSYtot*}, are obtained. 
For the finite temperature field theory case ($F^{1l}$), on the contrary, the integer $n$ and the $3D$ momentum $p$ are not intertwined. The $ d^3p$ integral provides the trace over quantum fluctuations (and is cut at a certain momentum cutoff), while the $\sum_n$ gives the statistical average (mixed states), and as such has to be performed all the way up to infinity to implement ergodicity. 
This infinite sum is at the origin of the finiteness\footnote{It is worth to note that the full result for $F^{1l}$ is finite only if the theory is supersymmetric.} of the T-dependent contribution $F_T^{1l}$  to $F^{1l}$. In fact, 
\begin{align}\label{common*}
	F_T^{1l} &= 
	\frac T2\sum_{n=-\infty}^\infty\int d^3 p \, \log(p^2+ m^2+ f_n)\nonumber \\
	& -  \frac12\int d^3 p \, \sqrt{p^2+m^2} \sim T^4 = {\rm finite}\,.
\end{align} 

As mentioned above and discussed at length in\,\cite{Branchina:2023ogv} and\,\cite{Branchina:2023rgi},  performing in the KK case too the sum over $n$ up to infinity (while keeping $p$ finite)  mistreats the loop momentum. This is what causes the (artificial) disappearance of the   $q$-dependent UV-sensitive terms
in\,\eqref{SUSYtot*}, giving the impression that there is a cancellation similar to what happens in the finite temperature case. 

Moving now to the example of the Casimir energy $\mathcal E_C$, we observe that the latter is nothing but the difference between the vacuum energy $\rho_{_R}$ calculated in the KK theory with one compact dimension (that for our illustrative purposes we took as a circle of radius $R$) and the corresponding $\rho_{_\infty}$ calculated in the  decompactification limit (i.e.\,\,in the theory with all non-compact dimensions): 
$\mathcal E_C = \rho_{_R} -\rho_{_\infty}$. Now, if we  perform the one-loop calculation of $\rho^{1l}$ according to previous literature,  meaning that in\,\eqref{common}  we perform  the sum over $n$ up to infinity, $\mathcal E_C^{1l}$ takes a form in all similar to\,\eqref{common*}, with $T^4$ replaced by $m_{_{\rm KK}}^4$. 
However, we have shown in\,\cite{Branchina:2023ogv} that (contrary to what is found in previous literature)  the  subtraction $\rho_{_R} -\rho_{_\infty}$ is not in general sufficient to ensure the disappearance of all the UV-sensitive terms. This is for instance the case of KK theories with Scherk-Schwarz SUSY breaking considered in the present work.  

\section{Summary, conclusions and open questions}

In summary,  we have shown that the criticisms raised by the authors of\,\cite{Anchordoqui:2023laz}  against our work\,\cite{Branchina:2023ogv} are unfounded. Their objections rely on Eqs.\,\eqref{domlambda*}, that in turn arise from the  replacement\,\eqref{replace*}, namely  $\rho^{1l}\to \Lambda_{cc}$,  operated by the authors of\,\cite{Anchordoqui:2023laz} in our original results for $\rho^{1l}$ (here reported in   Eqs.\,\eqref{domrho**}). We have shown  that \,\eqref{replace*}  is inconsistent and leads to arguments deprived of  physical meaning. The most striking consequence of the replacement\,\eqref{replace*} is that for the physical cutoff $\Lambda_{\rm SM}$ of the Standard Model it necessarily implies    $\Lambda_{\rm SM} \sim m_{_{\rm KK}} \sim 10^{-3} {\rm eV}$ (see Eq.\,\eqref{Lambda SM iden}), which is obviously absurd. This crucial consequence of their replacement, that makes their arguments inconsistent, was overlooked by the authors of\,\cite{Anchordoqui:2023laz}. Moreover, we have shown that, even closing an eye on this severe shortcoming, the arguments presented by the authors of\,\cite{Anchordoqui:2023laz} have further flaws and lack of internal coherence. 

Before ending this work, we would like to make a couple of additional comments on some open questions connected with the subject discussed in the present work. 

Concerning the UV-IR mixing invoked in\,\cite{Anchordoqui:2023laz},  there are several works where the impact that  modular invariance in string theory can have on its low energy EFT limit is studied (see for instance\,\cite{Feruglio:2017spp,Nilles:2020nnc,Abel:2021tyt,Abel:2023hkk}), and it is certainly of the  greatest importance to continue the investigation in these directions.  However, we think that such a delicate issue cannot be trivially shortcut with the choice  $\Lambda=m_{_{\rm KK}}$ considered  in\,\cite{Anchordoqui:2023laz} and discussed above.

Finally, we would like to spend one word on the general framework where, in our opinion, the DD scenario should be framed.  In\,\cite{Branchina:2023ogv} and in the present work we considered the contribution to the vacuum energy that comes  
from the bulk. The reason is that in the paper that contains the DD proposal\,\cite{Montero:2022prj}, as well as in the paper\,\cite{Anchordoqui:2023laz} that attempts to confute the results of our work\,\cite{Branchina:2023ogv}, this is the only part of the vacuum energy taken into account. Naturally, there is also a contribution from the Standard Model, that, as already said, lives on a 3-brane. Since the SM physical cutoff  $\Lambda_{\rm SM}$ is at least of the TeV order, this contribution cannot be ignored. In this respect, it is important to stress that it has been already objected in the past (well before the formulation of the DD proposal) that the SM contribution to the vacuum energy seems to invalidate any attempt to get a small cosmological constant in theories with compact extra dimensions\,\cite{Dvali:2000xg,Dvali:2002pe}.

The viability of the DD scenario is certainly subject to the  existence of a physical mechanism that could  dispose of this too large contribution. We expect that, for both bulk and brane  
contributions, such a mechanism (if any) should be provided by the piling up of quantum fluctuations in the UV $\to$ IR renormalization group (RG) flow that takes into account the whole $(4+1)$D theory, including the SM in the 3-brane. In such a framework, the UV-complete theory (string theory/quantum gravity) should provide the boundary (i.e.\,\,UV) value of the vacuum energy (more generally the boundary values of all the physical parameters), so that RG equations should dictate how the measured vacuum energy is reached in the IR (see\,\cite{Branchina:2022gll} for examples where this mechanism is realized in $d=3$ and $d=4$ scalar theories).
Work is in progress in this direction\,\cite{inprogress}.

\section*{Acknowledgments}
	We thank S. Abel, G. Dvali, H. P. Nilles, A. Pilaftsis and M. Scalisi for useful discussions.
	The work of CB has been partly supported by the Basic Science Research
	Program through the National Research Foundation of
	Korea (NRF) funded by the Ministry of Education, Science and Technology (NRF-2022R1A2C2003567), and partly by the European Union – Next Generation EU
	through the research grant number P2022Z4P4B “SOPHYA - Sustainable Optimised PHYsics
	Algorithms: fundamental physics to build an advanced society” under the program PRIN 2022
	PNRR of the Italian Ministero dell’Università e Ricerca (MUR).
	The work of VB, FC and AP is carried out within the INFN project  QGSKY. 

\appendix

\section{Dark Dimension scenario}
For the reader's convenience, in this Appendix we provide an expanded version of the introduction to the DD scenario  given in the text. The proposal is formulated 
combining observational data and phenomenological bounds with swampland conjectures.  It
is based on the assumption that our universe lies in a unique corner of the quantum gravity landscape. This region is determined by the 
distance  conjecture\,\cite{Ooguri:2006in}, according to which at large distances in the moduli field space $\phi$ the scale $\mu_{tow}$ of one of the towers present in these asymptotic regions  becomes exponentially small, $\mu_{tow} \sim e^{-\alpha |\phi|}$, with $\alpha$ positive $\mathcal O(1)$ constant. When this conjecture is applied to (A)dS vacua\footnote{The AdS version of the conjecture is supported by a larger amount of arguments in string theory.  The relation \eqref{distcon*} is nevertheless assumed to be valid also in dS space.}, the relation
\begin{equation}\label{distcon*}
	\mu_{tow}\sim\left|\frac{\Lambda_{cc}}{M_P^4}\right|^{\alpha}M_P
\end{equation} 
is established\,\cite{Lust:2019zwm}. 

Combining indications from  one-loop string calculations with the Higuchi bound\,\cite{Higuchi:1986py}, the authors of\,\cite{Montero:2022prj} restrict $\alpha$ to the range 
\begin{equation}\label{AdSconj*}
	\frac 14\leq \alpha\leq 
	\frac 12\,.
\end{equation} 
Moreover, according to the emergent string conjecture\,\cite{Lee:2019wij,Lee:2019xtm}, 
they observe that in these asymptotic regions only two  possible kinds of towers are expected: 
either a tower of string excitations, or one of KK states.

Since their goal is to apply these conjectures/results to our universe, they begin by observing that a light tower of states causes significant deviations from the gravitational Newton's
force law at the energy scale\,\footnote{For a KK tower this is due to gravity propagating on the extra dimensions, while for the string case local physics breaks down at the string scale.} $\mu_{tow}$, and that torsion
balance experiments provide the strongest bounds to deviations of the $1/r^2$ law. Noticing that this law was verified down to scales around\,\,30\,${\rm \mu m}$\,\cite{Lee:2020zjt}, which implies that $\mu_{tow}$ has to satisfy the bound
\begin{equation}
	\mu_{tow} \geq 6.6\,\, {\rm meV} ,
\end{equation}
and that  this value of $\mu_{tow}$ is close to the energy scale associated to the cosmological
constant, $\Lambda_{cc}^{1/4}\sim 2.31 \,\, {\rm meV}$, they conclude that 
$\mu_{tow}$ must satisfy the relation  
\begin{equation} \label{four}
	\mu_{tow}\geq \Lambda_{cc}^{\,\,1/4}\,,
\end{equation}
otherwise we would have already detected
deviations from Newton's law. They then observe that the only way to make the \vv experimental
bound'' \eqref{four} consistent with the \vv theoretical swampland bound''\,\eqref{distcon*} is to take  $\alpha=1/4$ (the quotation marks are to indicate that we are using their words). This leads them to conclude that:
(i) \vv there is a tower of states starting at the neutrino scale in our universe''; (ii) \vv this tower is either a light string tower or a light
KK tower''. Concerning this latter point, they observe that  \vv since we can describe physics above the neutrino scale with {\it Effective Field Theory}'' (our italics), the string scenario is \vv ruled out experimentally'', as no experiment has ever detected a string spectrum at these scales. 

The authors conclude that  experiments tell us that the light tower must be a KK tower, $\mu_{tow}=m_{_{\rm KK}}$, 
so that (in their words) \vv the only possibility left'' is that of 
a \vv scenario of  decompactification'', where \eqref{distcon*} becomes 
$\Lambda_{cc} 
\sim m_{_{\rm KK}}^4$, that is nothing but\,\eqref{CC-rel1}.
The last step in the construction of their proposal is the determination of the number of extra compact dimensions. To this end, they  compare bounds on $m_{_{\rm  KK}}$ that come from calculations of the  heating of neutron stars due to the surrounding cloud of trapped KK gravitons \cite{Hannestad:2003yd,PDG}  with the value of $m_{_{\rm KK}}$ given in \eqref{CC-rel1}, that leads to the conclusion that there can be  {\it only one} extra dimension of size $\sim \mu {\rm m}$.

Therefore,  assuming that our universe lies in a unique corner of the quantum gravity landscape, and  combining swampland conjectures with phenomenological and observational data, the Dark Dimension 
scenario predicts that at low energies physics is described by a $(4+1)$D EFT with a compact dimension of micrometer size. 

\section{Standard Model cutoff}

In this Appendix  we derive the  relation between the cutoff $\Lambda$ of a $(4+1)$D theory  and the $4$D cutoff $\Lambda_{\rm SM}$ of the Standard Model  localized on a 3-brane. Let us consider the $(4+1)$D theory (with compact space dimension in the shape of a circle of radius $R$) defined by 
\begin{equation}
	\mathcal{S}=\mathcal{S}_{\rm grav}+\mathcal{S}_{\rm{mat}},
	\label{free scalar action}
\end{equation}
where
\begin{equation}
	\label{action}
	\mathcal{S}_{\rm grav}=\frac{1}{2\hat{\kappa}^2}\int d^4xdz \sqrt{\hat{g}}\,\left(\hat{\mathcal R}-2\hat\Lambda_{cc}\right)
\end{equation}
is the $(4+1)$D Einstein-Hilbert action and as an example for the matter action we take 
\begin{equation}
	\label{scalar action}
	\mathcal{S}_{\rm mat}=\int d^4x dz\,\,\sqrt{\hat{g}}\left(\hat{g}^{MN}\partial_{M} \hat\Phi^* \partial_N \hat\Phi-m^2|\hat\Phi|^2\right),
\end{equation}
with $\hat \Phi$  a $(4+1)$D scalar field that obeys the boundary condition $\hat \Phi(x,z+2\pi R)=\hat \Phi(x,z)$. We indicate with $x$ the $4$D coordinates and with $z$ the coordinate along the compact dimension. Using the  signature $(+,-,-,-,-)$,
the $(4+1)$D metric is parametrized as
\begin{equation}
	\label{metric}
	\hat{g}_{_{MN}}=\begin{pmatrix}
		e^{2\alpha\phi}g_{\mu \nu}-e^{2\beta \phi}A_{\mu}A_{\nu} & e^{2\beta \phi}A_{\mu} \\ 
		e^{2\beta \phi}A_{\nu} & -e^{2\beta \phi}\,,
	\end{pmatrix}
\end{equation}
where $A_\mu$ is the so called graviphoton and $\phi$ the radion field. 
Considering only zero modes for $\hat g_{_{MN}}$, i.e.\, $g_{\mu\nu}(x)$, $A_\mu(x)$ and $\phi(x)$
only depend on $x$, and  integrating over $z$, for the 4D gravitational action $\mathcal{S}^{(4)}_{\rm grav}$ we get\,\cite{Benakli:2022shq}
\begin{align}
	\label{grav4D}
	&\mathcal{S}^{(4)}_{\rm grav}=\frac{1}{2\kappa^2}\int \mathrm{d}^4x\,\sqrt{-g}\,\,\left[ \mathcal R-2e^{2\alpha\phi}\hat\Lambda_{cc}+{2}\alpha \Box \phi+\frac{(\partial \phi)^2}{2}-\frac{e^{-6\alpha\phi}}{4}F^2\right], 
\end{align}
where the $4$D constant $\kappa=M_P^2$ is related to the $(4+1)$D $\hat{\kappa}=\hat M_P^3$ through the relation $\kappa^2=\hat{\kappa}^2/(2\pi R)$. The fields 
$\phi$ and $A_\mu$ in the above equation are dimensionless (dimensionful fields are obtained through the redefinition $\phi\to\phi/(\sqrt{2}\kappa),\, A_{\mu}\to A_{\mu}/(\sqrt{2}\kappa)$), and we used $2\alpha+\beta=0$. The canonical kinetic term in \eqref{grav4D} for the radion field is obtained taking  $\alpha={1}/{\sqrt{12}}$.

Considering the Fourier decomposition of $\hat\Phi(x,z)$, for the $4D$ matter action \eqref{scalar action} we have 
\begin{align}
	\label{mat4D}
	&\mathcal S^{(4)}_{\rm mat}=\int d^4x \sqrt{-g}\,\,\sum_{n}\left[\left|D\varphi_{n}\right|^2-\left(e^{\sqrt{\frac{2}{3}}\frac{\phi}{M_P}} m^2+e^{\sqrt 6\frac{\phi}{M_P}}\frac{n^2}{R^2}\right)\left|\varphi{_n}\right|^2 \right],
\end{align}
where  
$D_\mu\equiv \partial_\mu-i\left({n}/{R}\right)A_\mu$,
and $\varphi_n(x)$ are the KK modes of $\hat \Phi(x,z)$. 

Taking a constant background for the radion (that for notational simplicity we continue to indicate with $\phi$) and the trivial background for $A_\mu$, the metric \eqref{metric} becomes
\begin{equation}
	\label{bmetric}
	\hat{g}^0_{_{MN}}=\begin{pmatrix}
		e^{\sqrt{\frac{2}{3}}\frac{\phi}{M_P}}\eta_{\mu \nu} & 0 \\ 
		0 & -e^{-2\sqrt{\frac{2}{3}} \frac{\phi}{M_P}}
	\end{pmatrix}.
\end{equation}
From \eqref{mat4D} we define  the $\phi$-dependent radius $R_\phi\equiv R\, e^{-\sqrt{\frac32}\frac{\phi}{M_P}}$. With such a definition, we immediately see that, when computing radiative corrections, the $(4+1)$D momentum $\hat p\equiv (p,n/R)$ is cut as
\begin{equation}
	\hat p^2=e^{-\sqrt{\frac23}\frac{\phi}{M_P}}\left(p^2+
	\frac{n^2}{R_\phi^2}\right)\le \Lambda^2.  
\end{equation}
This latter equation is conveniently rewritten as
\begin{equation}\label{cutoff4}
	p^2+
	\frac{n^2}{R_\phi^2}\le \Lambda_\phi^2,
\end{equation} 
where we defined $\Lambda_\phi\equiv \Lambda \,e^{\frac{1}{\sqrt 6}\frac{\phi}{M_P}}$. In terms of the dimensionless $\phi$ of \eqref{metric} and \eqref{grav4D}, and before using $\alpha=1/\sqrt{12}$, it is $\Lambda_\phi=e^{\alpha\phi}\Lambda=m_{_{\rm KK}}^{1/3}R^{1/3}\Lambda$.

As $p^2$ in\,\eqref{cutoff4}  is nothing but the modulus of the four-momentum on the brane,  this equation tells us that $\Lambda_\phi$ is the cutoff $\Lambda_{\rm SM}$ of the SM (or more generally of the BSM model that lives on the 3-brane, where fields have $n=0$). Therefore:
\begin{equation}\label{cutoff5} \Lambda_{\rm SM}=\Lambda_\phi = \Lambda \,e^{\frac{1}{\sqrt 6}\frac{\phi}{M_P}}. 
\end{equation} 
Finally, as the DD scenario is realized for negative values of $\phi$, from \eqref{cutoff5} we see that $\Lambda_{\rm SM} \leq\Lambda $, i.e.\, the SM cutoff is lower than the cutoff of the $(4+1)$-dimensional EFT that implements the DD scenario. 

Before closing this Appendix, let us note that here we considered a spherical cutoff. Naturally, we can make a different choice, taking for instance a cylindrical cutoff \cite{Branchina:2023rgi}
\begin{equation*}
	p^2\le \Lambda_\phi^2 \qquad {\rm and} \qquad \frac{n^2}{R_\phi^2}\le \Lambda_\phi^2.
\end{equation*}
This choice, that is  closer to what is typically done when using the species scale $\Lambda_{\rm sp}$ as the cutoff \cite{Grimm:2018ohb}, does not change the above considerations.

\end{document}